\documentclass[%
 reprint,
 amsmath,amssymb,
 aps,prd,
]{revtex4-2}

\usepackage{graphicx}
\usepackage{dcolumn}
\usepackage{bm}
\usepackage[mathlines]{lineno}
\usepackage{xcolor}

\begin{document}


\title{Ultrahigh-energy cosmogenic neutrino emissions in the high-redshift universe}

\author{Shigeru Yoshida}
\email{syoshida@hepburn.s.chiba-u.ac.jp}
\author{Maximilian Meier}
\thanks{maximilian.meier@icecube.wisc.edu}
\affiliation{%
 International Center for Hadron Astrophysics, Chiba University, Chiba 263-8522, Japan\\
}%


\date{\today}

\begin{abstract}
The James Webb Space Telescope (JWST) revealed a large population of active galactic nuclei (AGN) with redshifts greater than five. We show that if they emit ultrahigh-energy protons with energies up to $\lesssim 10^{19}$~eV, the cosmogenic neutrino production in the high-redshift CMB field yields a neutrino flux with a bump at around 50~PeV. This flux is potentially consistent with the current neutrino intensity estimate by the IceCube Neutrino Observatory within the parameter space allowed from the AGN luminosity and number density estimates by JWST. Future neutrino observations that confirm the 50-PeV bump and constrain the small-scale anisotropy will infer ultra-high energy cosmic-ray emissions in the early universe. 
\end{abstract}

\maketitle

High-energy neutrino observations provide a novel probe for understanding cosmic-ray origins. Secondary neutrinos resulting from interactions with ultrahigh-energy cosmic rays (UHECRs) with energies far beyond EeV ($10^{18}~{\rm eV}$) hold promise as probes of UHECR origins, including source directions, emission timescales, and relevant energetics. This potential is attributable to the inherent property of neutrinos, namely their ability to traverse cosmological distances while retaining their primary energies, with the exception of losses incurred through the cosmological adiabatic process~\cite{Seckel:2005cm}. The IceCube Neutrino Observatory has measured the high-energy neutrino flux in the energy region beyond PeV~\cite{Aartsen:2013bka, Aartsen:2016ngq}. The detection of multiple PeV-energy neutrino events has been achieved~\cite{Aartsen:2016xlq, Aartsen:2018vtx, IceCube:2021rpz,IceCubeCollaborationSS:GlobalFit2026}, thereby facilitating the estimation of the energy flux to be approximately $\mathcal{O}(10^{-9})~{\rm GeV/(cm^2 sec\ sr)}$. However, the origin of these neutrinos remains to be elucidated.  Are these neutrinos definitely associated with UHECR emissions? Does the picture of UHECR and PeV-energy neutrino connections line up with the null detection of further higher-energy neutrinos by IceCube~\cite{Aartsen:2018vtx, IceCubeCollaborationSS:2025jbi} and Pierre Auger Observatory~\cite{Aab:2019auo,PierreAuger:2023pjg}? These fundamental questions still need to be answered to solve the long-standing mysteries of UHECR production mechanisms.

An interesting discovery in this context has been recently made by the James Webb Space Telescope (JWST). It surprisingly revealed a large population of active galactic nuclei (AGN) in the far-universe~\cite{Onoue:2022goe, Kocevski:2023,2023ApJ...959...39H}. Their bolometric luminosity in the infrared band is between $10^{43-46}$~erg/s distributed at redshifts from $z\sim5$ to 10~\cite{deGraaff:2025uvk, Umeda:2025bha}.  These AGNs are called 'Little Red Dots' (LRDs), reflecting the observational fact that they appear as compact red objects~\cite{Matthee:2023utn, Greene:2024phl, Kocevski:2025tft, Labbe:2025, Perez-Gonzalez:2024dul}. Nearly $\lesssim 20$~\% of the faint AGNs detected by JWST have been classified as LRDs exhibiting distinct features of a V-shape spectral energy distribution (SED), with blue UV and red optical continuum~\cite{2025A&A...698A.227F, Setton:2025}. The fraction of LRDs in the observed AGN sample is found to be increasing with redshift, indicating that they are among representative energetic compact objects in the high-$z$ universe~\cite{2025ApJ...995...21T, Taylor:2024}. The estimated comoving number density reaches $\lesssim 10^{-4}~{\rm Mpc}^{-3}$~\cite{Kocevski:2025tft, Umeda:2025bha}.

The nature of LRDs is still largely unknown, mostly because they exhibit peculiar properties such as their V-shape SEDs, very faint X-ray emissions, and weak variability of emissions~\cite{Kokubo:2024ukw, Yue:2024yti, Ananna:2024jug, Maiolino:2024uon}, all of which are not common in normal AGNs. Their nature is an open question, which requires further observational and theoretical studies. Nevertheless, their luminous emissions suggest that they may be powerful enough to accelerate protons to ultrahigh energies. The dimension analysis, known as the Hillas condition, indicates that the maximal accelerated energy of protons is given by ${\varepsilon_p^{\rm max}\simeq 8\times 10^{10}(L_\gamma/(10^{45}~{\rm erg/s}))^{1/2}\xi_{\rm B}^{1/2}}$~GeV, where $\xi_{\rm B}$ is the equipartition parameter of the B-field with respect to the photon luminosity~\cite{Lemoine:2009pw}. It is not implausible to consider that LRDs may be UHECR sources.
Since the metallicity is naturally expected to be low in $z\gtrsim 5$, which has also been suggested by the very weak {\sc [Oiii]} lines in the host-galaxies~\cite{Maiolino:2025tih},  it is a reasonable assumption that they emit protons rather than heavier nuclei.

\begin{figure*}[t]
\includegraphics[width = 0.85\textwidth]{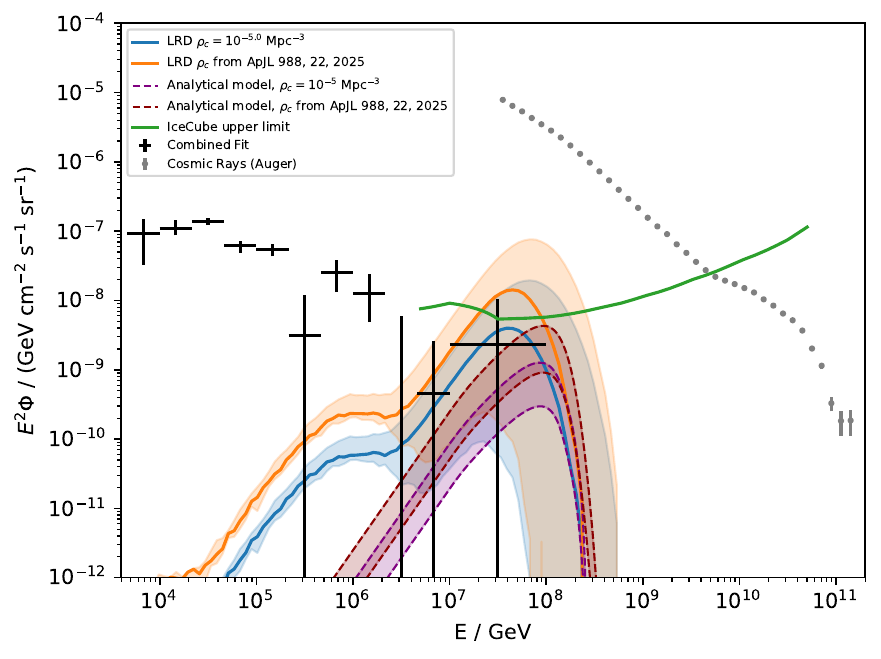}
\caption{\label{fig:Neutrino_flux} The all-flavor cosmogenic spectrum of neutrinos generated by ultrahigh-energy protons in the early universe. Full mixing in the neutrino flavor oscillation during the propagation is taken into account. $L_{\rm CR}=10^{45}~{\rm erg/s}$ is assumed. The blue and orange lines have been obtained by \textsc{crp}{\small{ropa}} simulations with a maximum proton energy of $\varepsilon_p^{\mathrm{max}} = 10^{10} \, \mathrm{GeV}$, while the shaded bands indicate the effect of varying $\varepsilon_p^{\mathrm{max}}$ by a factor of $2$. The black data points with crosses were obtained by the joint analysis using the tracks and cascade samples measured by IceCube~\cite{IceCubeCollaborationSS:GlobalFit2026}. The thick line indicates the differential upper limit obtained by the extremely high-energy (EHE) neutrino search~\cite{IceCubeCollaborationSS:2025jbi}. The bands delineated by the dashed curves represent the estimations derived from the semi-analytical formula. The width of the band is indicative of the uncertainty associated with the rectangular integration method. Two cases of the cosmological evolutions are presented: No evolutions with a constant comoving number density of $10^{-5}~{\rm Mpc}^{-3}$, and the log-normal distribution model proposed in Ref.~\cite{Inayoshi:2025isg}. The small gray data points represent UHECR flux measured by Pierre Auger Observatory~\cite{Fenu:2017hlc}. Note that protons emitted in the early universe do not reach Earth and are not associated to the UHECR flux data.}
\end{figure*}
If LRDs are emitting protons with energies beyond EeV, these protons inevitably collide with the cosmic microwave background (CMB) photons. The channel of the photohadronic interactions with CMB photons opens when $\varepsilon_p\gtrsim 10^{10}((1+z)/8)^{-1}$~GeV. The average interaction length
is only $\lambda_{p\gamma} \lesssim  400 ((1+z)/8)^{-3}$~kpc at this energy and above, with the intergalactic medium being effectively opaque to protons with energies beyond the photohadronic reaction threshold energy. On the contrary, the secondary neutrinos resulting from the interactions in the dense CMB field penetrate the high-redshift universe to reach our galaxy. They are ultrahigh-energy messengers to probe UHECR production in the early universe. The production of cosmogenic neutrinos~\cite{Beresinsky:1969qj}, which is often discussed in the context of UHECRs propagating in a lower redshift universe in the literature~\cite{Yoshida:1993pt, Engel:2001hd, Ahlers:2010fw, Kotera:2010yn, Decerprit:2011qe, Yoshida:2012gf, Aloisio:2015ega}, can provide the very efficient neutrino generation mechanism that must occur in a high-$z$ space filled with UHECR emitters.

{\it Flux of ultrahigh-energy neutrinos--}
We calculate the flux of neutrinos produced during the UHECR proton propagation in the high-redshift universe using the \textsc{crp}{\small{ropa}} simulation package~\cite{CRPropa:2022ovg}. The injected protons are traced from their sources to Earth, accounting for all the relevant interaction processes, such as photopion production and pair production, in the CMB and extragalactic background light (EBL) fields. The neutron decay process and the redshift adiabatic loss are also taken into account. The secondary neutrinos produced during the propagation of protons are summed in the redshift space considering the $\Lambda$CDM cosmology. The primary proton spectrum is assumed to be a power-law spectrum $d\dot{N}_p/d\varepsilon_p \sim \varepsilon_p^{-2}$ from 10 PeV to 10 EeV. The maximum energy value $\varepsilon_p^{\rm max} $ is conservatively chosen to be well below the Hillas limit. The bolometric UHECR proton luminosity is assumed to be $L_{\rm CR}=10^{45}\ {\rm erg/s}$, which is comparable to the near-infrared bolometric photon luminosity of LRDs observed by JWST. The JWST data indicate that most LRDs are found from $z=5$ to 10 without strong features of cosmological evolution. In our simulation, we assume that the proton emitters are evenly distributed within the comoving volume from $z=5$ to 10, with no evolution. This assumption is conservative with regard to neutrino production. The comoving number density of LRDs estimated from JWST observations is still statistically uncertain, and we take the face value of $n_0=10^{-5}~{\rm Mpc^{-3}}$~\cite{Akins:2025iqo, Kokorev:2024kqm, Umeda:2025bha}, but we will consider the other cases later.

Figure~\ref{fig:Neutrino_flux} shows the calculated all-flavor neutrino fluxes. The data points representing IceCube measurements of the all-sky astrophysical neutrino flux~\cite{IceCubeCollaborationSS:GlobalFit2026} are also plotted for comparison. The neutrino spectrum has a pronounced bump at $\sim 50$~PeV. This bump feature is created based on the two effects: the $\Delta$(1232)-resonance effect in the photopion production, and the effect of the energy threshold to make the mass of a pi-meson in a collision. While the conventional cosmogenic neutrino energy flux has a bump at around EeV, a much higher temperature of the CMB in the early universe, combined with the large adiabatic energy loss, substantially lowers the bump energy. It is notable that the peak intensity at the bump can explain the IceCube data above 10~PeV. 

The neutrino spectrum obtained by the simulations employing \textsc{crp}{\small{ropa}} has been investigated by a semi-analytical calculation. Photohadronic interactions in the high-$z$ CMB field result in rapid energy losses of cosmic-ray protons, leading to quasi-calorimetric conversions of energies carried by protons into those carried by the secondary particles.  This circumstance enables the estimation of the resulting neutrino flux by adopting the formulation for the neutrino emission from a source via $p\gamma$ interactions with adjustments to adequately describe the evolution of the CMB field. The formulation is described in the {\it Supplemental Material}~\cite{suppl}~(see also references~\cite{Yoshida:2024fiu, Yoshida:2014uka, Mucke:1999yb} therein).  The band delineated by the dashed curves in Fig.~\ref{fig:Neutrino_flux} indicates the analytical estimate. The spectral shape is narrower due to the approximation that neglects the Bethe-Heitler pair creation process and the contributions from interactions with EBL photons. However, the bump structure and its peak intensity are consistent with the results obtained by the full numerical simulation within the uncertainty arising from the simple analytical approximations. For further elaboration, refer to the follow-up discussions in the {\it Supplemental Material}~\cite{suppl}. We conclude that the prediction of the primary feature of the neutrino spectrum is sound and solid based on basic principles of physics in the CMB field.

If LRDs emit nuclei instead of protons, the resulting cosmogenic flux is substantially lower than what is presented here. Photodisintegration interactions in the CMB field break up the nuclei, and cosmogenic neutrino generation is mostly driven by the limited fraction of secondary nucleons whose energies are above the photohadronic energy threshold. We calculated the flux of neutrinos for the case of helium emission by \textsc{crp}{\small{ropa}}. We set $\varepsilon_{\rm He}^{\rm max}=2\times 10^{10}$~GeV, which corresponds to the circumstances of accelerating protons to $\simeq 10^{10}$~GeV. This calculation reveals that the resultant flux is two orders of magnitude lower. Proton-dominated emission is an unavoidable condition for a plausible flux in the present model.


{\it Discussion--}
The predicted intensity of neutrinos that is consistent with the IceCube measurement is derived from the representative parameter values obtained through the JWST observations of LRDs. That is, the photon luminosity ($10^{45}$~erg/s), the comoving number density ($10^{-5}\ {\rm Mpc^{-3}}$), and the range of their emission epochs in the redshift space ($5 \leq z \leq 10$). 
The most sensitive variables to the intensity of the neutrino flux, which remain poorly or not constrained by observations, are the cosmological evolution parameter and the maximum energy of protons emitted from an LRD. The following discussion will address these issues.

The present JWST data cannot statistically exclude a weak evolution. The assumption of the absence of evolution results in a conservative prediction of the flux intensity. Employing the weak log-normal evolution model of the comoving number density of the LRDs, which has been proposed for interpreting the JWST data in the context of the AGN multi-episode scenario~\cite{Inayoshi:2025isg}, results in an approximately factor-of-three increase in flux, as illustrated in Fig.~\ref{fig:Neutrino_flux}. 

The adopted maximum proton energy is not firmly rooted in observational evidence. The X-ray emission observed in LRDs is typically faint, suggesting that the system is optically thick and may be enveloped by a cocoon-like photon wall~\cite{Naidu:2025rpo, deGraaff:2025uvk, Rusakov:2026}. In this environment, the acceleration of protons to $\sim$~10~EeV appears improbable, although it aligns with the Hillas criterion. In this scenario, the maximum proton energy would be below PeV. As a result, LRDs would appear as hidden neutrino sources in the 10 TeV region~\cite{Kuze:2026jrm}. However, the cocoon hypothesis is far from being established. A thick gas environment would naturally drive bright infrared radiation at $\lambda\gtrsim 10\mu{\rm m}$, due to the reheating of dust. However, this emission has not been detected~\cite{Perez-Gonzalez:2024dul, Casey:2025yfl}.  The broad Balmer emission lines observed in the JWST data imply a clumpy gas structure (e.g. Refs.~\cite{Tang:2026, Ji:2026}), contradicting the hypothesis of a completely closed, cocoon-like environment. Beyond LRDs, the more commonly observed population of broad-line AGN~\cite{2023ApJ...959...39H} may also be capable of accelerating protons up to 10 EeV.
 While their estimated bolometric photon luminosities are typically fainter compared to those of LRDs, with values around $\sim 3\times 10^{44}~{\rm erg/s}$, the population of these sources ($n_0\sim 2\times 10^{-4}~{\rm Mpc}^{-3}$) is approximately an order of magnitude greater than that of LRDs in the same redshift region~\cite{2025ApJ...991...74G}. It is conceivable that the proton emission from a subset of these abundant AGNs might generate a comparable flux of neutrinos. Additional observational evidence is required to definitively rule out the UHECR scenario. Conversely, the existence of the 50-PeV neutrino bump can serve to suggest or exclude the UHECR model. Furthermore, the peak intensity is capable of constraining the maximum proton energy range, as the intensity is highly sensitive to the maximum energy. As illustrated in Fig.~\ref{fig:Neutrino_flux}, a factor-of-two variation in the maximum proton energy results in an order-of-magnitude difference in the peak intensity. This is because the amount of the cosmogenic neutrino production in the high-redshift universe is a consequence of the photohadronic interaction energy threshold profile determined by the evolving CMB temperatures with redshift and the cosmological epoch when the maximum proton energy exceeds the threshold. 

\begin{figure}[t]
\includegraphics[width=0.49\textwidth]{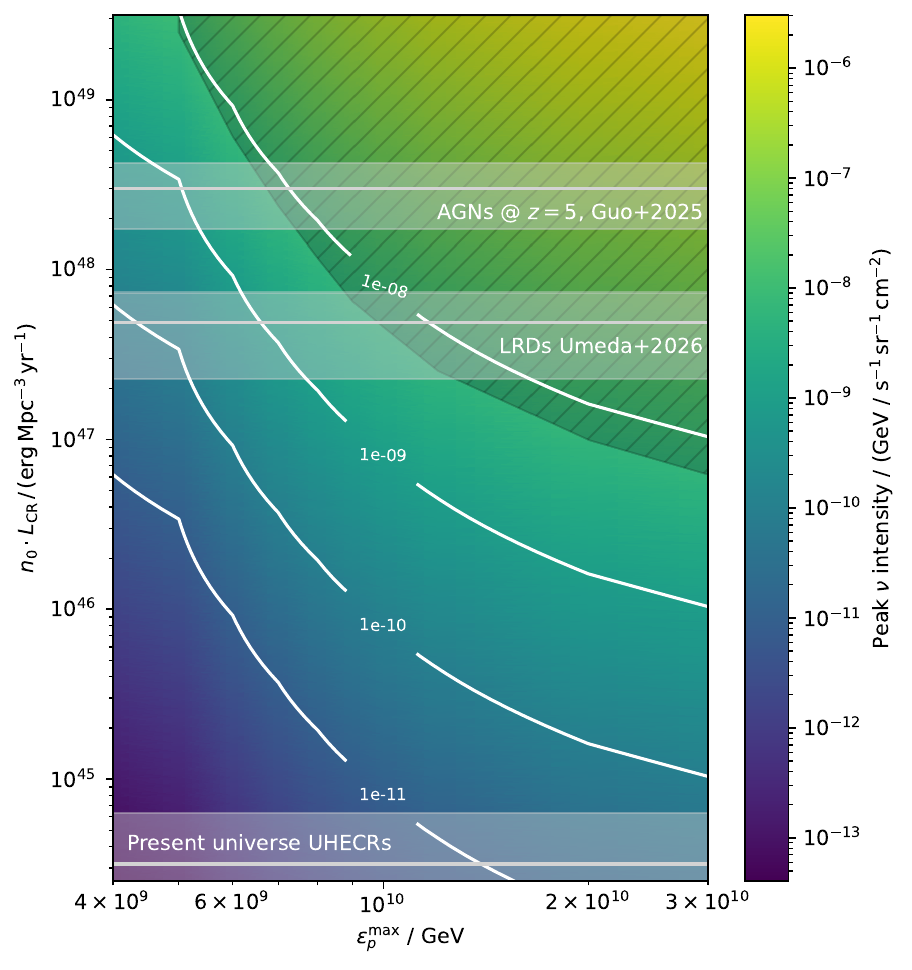}
\caption{The peak intensity of the neutrino energy flux in the plane of maximum proton energy $\varepsilon_p^{\rm max}$ and the UHECR luminosity density $n_0L_{\rm CR}$. The three bands represent the luminosity density range for the near-infrared photon emission from the broad-line AGNs~\cite{2025ApJ...991...74G}, LRDs~\cite{Umeda:2025bha}, and the emission of UHECRs in the present universe~\cite{Murase:2018utn}, respectively, taking into account the statistical uncertainties of the observational data.
The values of broad-line AGNs and LRDs presented in the plot are the comoving-volume-averaged values, which provide a reference indication based on the measurements available in limited redshift ranges at $z\gtrsim 5$ (extrapolated from $z=5$~\cite{2025ApJ...991...74G} for the former,  measured at $z\sim 6$ and $z\sim 8$~\cite{Umeda:2025bha} for the latter). The area filled with diagonal lines represents the parameter space rejected by the extremely-high energy (EHE) neutrino flux upper limit placed by IceCube~\cite{IceCubeCollaborationSS:2025jbi}. 
 \label{fig:flux_parameter_space}}
\end{figure}

Figure~\ref{fig:flux_parameter_space} illustrates the dependence of the resultant flux on the maximum proton energy and the UHECR proton luminosity density in the comoving space. The parameter space in which the flux of neutrinos is $E_\nu^2\phi\gtrsim 10^{-9} \, {\rm GeV/(cm^2 \, sec \, sr)}$ but compatible with the IceCube EHE limit~\cite{IceCubeCollaborationSS:2025jbi} is consistent with the luminosity density of the LRDs, provided $\varepsilon_p^{\rm max} \sim 5 \times 10^9 - 10^{10}~{\rm GeV}$.  For the more abundant, broader-line AGN population, channeling approximately $10-30\%$ of their energy budget into UHECR proton emission could yield a plausible flux in the present model. The indicated UHECR luminosity density is approximately two to three orders of magnitude higher than that estimated by the observation of UHECRs in the present universe~\cite{Murase:2018utn}. We note that the UHECR density can differ substantially in the high-$z$ universe, since the AGN comoving number density in high-redshift space is 50 times larger than the estimate that is simply extrapolated from quasars~\cite{2023ApJ...959...39H}. The $\varepsilon_p^{\rm max}$ dependence is only slightly pronounced below $\sim 4\times 10^{9}~{\rm GeV}$ because UHECR protons do not nearly interact with CMB photons but only with EBL photons in this case.

\begin{figure}[t]
\vspace{-2.8em}
\includegraphics[width=0.43\textwidth]{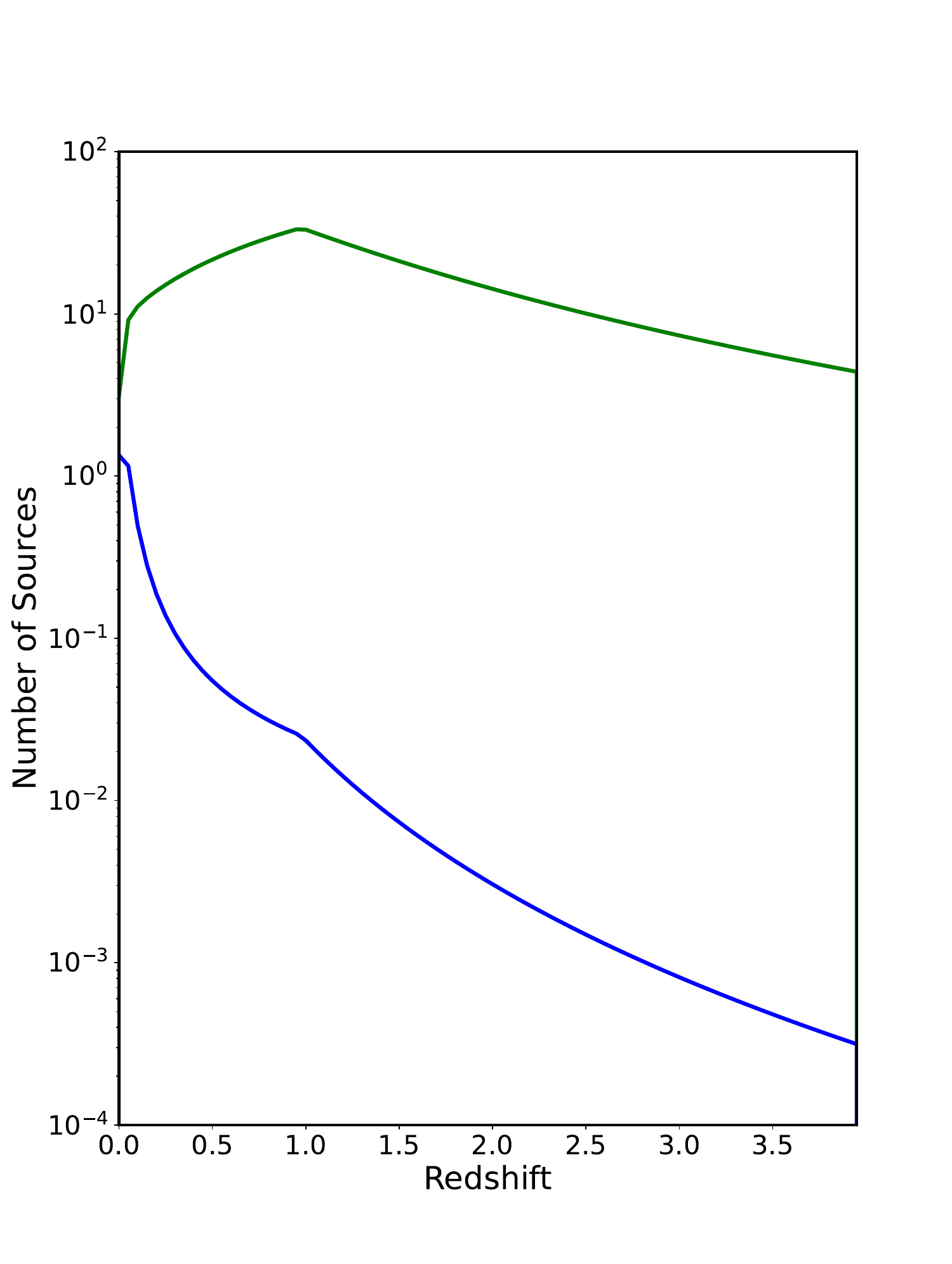}
\caption{\label{fig:multiplet} Distribution of neutrino sources per redshift bin ${\Delta z = 0.05}$ observed in the $2\pi$ sky to result in one detected neutrino event (green) and multiple events (blue) with an observation with an exposure of $1.3\times 10^{17}\ {\rm cm^2\ sec}$, if the sources are mostly in lower-redshift universe, following the star-formation history, in contrast to the present LRD model. The case of $L_{\rm CR}=10^{45}~{\rm erg/s}$, $n_0=3\times 10^{-7}~{\rm Mpc^{-3}}$ is presented for illustrative purposes. The cosmic evolution tracing the star-formation rate is assumed.}
\end{figure}

The $50~\rm{PeV}$ bump is a distinctive feature to suggest the connection between AGNs and UHECRs in the high-redshift universe. A bump structure is a common feature in the neutrino sources emitted via the $p\gamma$ reaction, a process that is frequently employed in the context of neutrino production at normal AGNs. However, the majority of the predicted bump energies is below 10~PeV (e.g. Refs.~\cite{Dermer:2014vaa, Kimura:2020srn, Kimura:2020thg}), particularly in the models involving an accretion disk. The reasons are twofold. A magnetic field of a certain strength, which is commonly required for UHECR acceleration, inevitably causes synchrotron cooling of muons (and pions). This process effectively suppresses the flux of neutrinos with energies well exceeding 10~PeV (see also Ref.~\cite{Yoshida:2020div} for the relevant discussions). Moreover, a significant number of AGNs do not provide an adequate supply of long-wavelength infrared ($\sim 0.1$~eV) photons, which would serve as the collision target for protons to efficiently produce neutrinos at energies ranging from 30 to 100~PeV.  In contrast, a distinct bump well above 10~PeV is a natural consequence of the cosmogenic neutrino generation in a high-redshift universe. We note that the blazar jet model (e.g.~\cite{Murase:2014foa, Padovani:2015mba}) could form a moderate bump structure at energies beyond 100~PeV. However, the predicted well-extended spectral shape has begun to impose severe constraints on the jet model due to the upper limit on EHE flux placed by IceCube~\cite{Aartsen:2016ngq,IceCubeCollaborationSS:2025jbi}.

A high-statistics measurement by a future giant neutrino telescope would be able to further test the present scenario. If a majority of steady sources contributing to the flux of neutrinos with energies beyond PeV are in a relatively nearby universe ($z\lesssim 1$), it is possible for multiple neutrinos to be detected from the same single source. For illustrative purposes, we assume hypothetical $p\gamma$ emitted sources of $L_{\rm CR}=10^{45}~{\rm erg/s}$ and comoving number density of $3\times 10^{-7}~{\rm Mpc^{-3}}$ following the star-formation compatible evolution from $z=0$ to 4. The parameters are within the ranges frequently discussed in the models of neutrino emission from the hot disk in radio-loud AGNs (e.g. Ref.~\cite{Kimura:2020srn}) and configured in a manner that would yield a comparable all-sky neutrino flux intensity shown in Fig.~\ref{fig:Neutrino_flux}.  Figure~\ref{fig:multiplet} shows the redshift distribution of neutrino sources that would result in event detections by an observation with an exposure of $1.3\times 10^{17}\ {\rm cm^2\ sec}$, which is equivalent to a 20-year observation by a neutrino telescope that is 50 times larger than the detection area of IceCube. The calculation method is described in Ref.~\cite{Yoshida:2022idr}. It indicates that $4\sim 5$ nearby sources ($z < 1$) yield detectable multiple neutrino events, thereby providing an identifiable signature of the small-scale anisotropy. If all the sources are distant in a high-redshift space, as in the case with the present hypothesis, it is entirely impossible to detect multiple events from a single source.

{\it Summary--}
If the recently discovered population of AGNs in the high-redshift universe emits protons with energies up to $\lesssim 10^{19}$~eV, the inevitable cosmogenic neutrino production in the cosmologically evolving CMB field could result in a sizable neutrino energy flux with a distinctive bump at around 50~PeV.  Future confirmations of this bump by next-generation neutrino observatories will offer the initial examination of the hypothesis of UHECR emissions in the early universe.

{\it Acknowledgments--}
The authors are grateful to Yuichi Harikane and Shigeo Kimura for useful discussions. S.Y.~acknowledges the Yukawa International Seminar for the year 2026a for productive discussions about the recent discoveries made by JWST. This work is supported by JSPS KAKENHI Grants No. 23H04892, and Institute for Advanced Academic Research of Chiba University.
\bigskip

\appendix
\onecolumngrid

\begin{center}
    {\bf Supplemental Material\\ Analytical Formulas for calculating the cosmogenic neutrino flux in a high-redshift universe}
\end{center}

The rapid energy cooling of protons due to the dense CMB field in a high-redshift universe enables the application of the modeling of neutrino sources emitted via the photohadronic interactions in order to adequately describe the cosmogenic neutrino production. Here we outline a set of semi-analytical equations to calculate the cosmogenic neutrino flux based on the generic model of the neutrino source emissions in Ref.~\cite{Yoshida:2024fiu}.

The average number of $p\gamma$ collisions at the $\Delta$(1232)-resonance energy when a proton travels the distance of $R$ is given by $\tau^0\equiv Rn_{\rm CMB}\sigma^\Delta_{p\gamma}$, where $n_{\rm CMB}$ is the CMB number density and $\sigma^\Delta_{p\gamma}$ is the cross section of the charged pion production at the $\Delta$-resonance energy. 
When the injection spectrum of protons from a source is written as $d\dot{N}/d\varepsilon_p=\kappa_{\rm CR}/\varepsilon_p^0 (\varepsilon_p/\varepsilon_p^0)^{-\alpha_p}$,
the generic model of the $p\gamma$ emission provides the following approximate formula for the differential flux of neutrinos from a UHECR source at redshift $z$ (e.g. Refs.~\cite{Yoshida:2012gf, Yoshida:2014uka}):
\begin{equation}
    \frac{d\dot{N}_\nu}{d\varepsilon_{\nu}}\approx 
\frac{L_{\rm CR}}{(\varepsilon_p^0)^2 \mathcal{R}}
\frac{3}{1-r_\pi}\frac{1}{x_\Delta^+-x_\Delta^-} 
\times \int_{\frac{\varepsilon_\nu}{(1-r_\pi)x_\Delta^+}}^{\varepsilon_p^{\rm max}} d\varepsilon_p\frac{\tau_{p\gamma}[\varepsilon_p, z]}{\varepsilon_p}\left(\frac{\varepsilon_p}{\varepsilon_p^0}\right)^{-\alpha_p}\ln{\left(\frac{x_\Delta^+}{\xi_p}\right)},
\label{eq:neutrino_yield_thermal}
\end{equation}
where $r_\pi\equiv m_\mu^2/m_\pi^2\simeq 0.57$ and
\begin{equation}
    \tau_{p\gamma}[\varepsilon_p, z] = \tau^0 \frac{\Delta_s(s_\Delta-m_p^2)}{16\zeta_3 (k_{\rm B}T_{\rm CMB}(1+z))^2\varepsilon_p^2} \\
    \times \{ -\ln{(1-e^{-\frac{s_\Delta-m_p^2}{4k_{\rm B}T_{\rm CMB}(1+z)\varepsilon_p}})}\}.
    \label{eq:optical_depth_proton}
\end{equation}
Here $s_\Delta$ is the Mandelstam variable at the $\Delta$(1232)-resonance of 
the photomeson interactions whose width is $\Delta_s$, $\mathcal{R}$ is the conversion factor from the bolometric luminosity $L_{\rm CR}$ to the spectral normalization factor $\kappa_{\rm CR}$,
$\zeta_3\simeq 1.202$ is the Riemann zeta function value at $3$, and 
$x_\Delta^{+(-)}$ is the maximal (minimal) bound of the relative energy of the emitted pion
normalized by the parent cosmic-ray energy. A kinematic relation represents them as
\begin{eqnarray}
x_{\Delta}^{\pm} &=& \overline{x_{\Delta}} \pm \frac{\sqrt{(s_\Delta + m_{\pi}^2 - m_p^2)^2 - 4 s_\Delta m_{\pi}^2}}{2 s_\Delta}, \nonumber\\
\overline{x_{\Delta}} &=& \frac{(s_\Delta + m_{\pi}^2 - m_p^2)}{2 s_\Delta}. 
\end{eqnarray}
and the coefficient derived from the inelasticity of the collision, $\xi_p$, is given by
\begin{equation}
\xi_p = \begin{cases}
x_\Delta^-                         & \varepsilon_\nu \leq (1-r_\pi)x_\Delta^{-} \varepsilon_p,\\
{\varepsilon_\nu\over (1-r_\pi)\varepsilon_p} & \text{otherwise}.
\end{cases}
\label{eq:xi_range}
\end{equation}
We set $\tau^0=1/\overline{x_\Delta}\simeq 4.8 $ by meeting the quasi-calorimetric condition.

The all-sky flux from these sources across the universe is then given by
\begin{equation}
\Phi_\nu(E_\nu)=\frac{cn_0}{4 \pi}
\int\limits_{z_{\rm min}}^{z_{\rm max}}dz\psi[z](1+z) \left|\frac{dt}{dz}\right|
\frac{d\dot{N}_\nu}{d\varepsilon_\nu}\left|_{\varepsilon_\nu=E_\nu(1+z)} \right.
\label{eq:diffuse_flux}
\end{equation}
where $n_0$ is the comoving source number density and $\psi[z]$ represents the cosmic evolution of the spectral emission rate per comoving volume. Our baseline assumption of the no-evolution case corresponds to $\psi[z]=1$. When implementing the log-normal evolution model, $n_0\psi[z]$ is given by the parametrization in Ref.~\cite{Inayoshi:2025isg}.

The above formulations were confirmed to provide compatible results given by the more analytical and less numerical formulation~\cite{Yoshida:2012gf} which is valid when the maximal energy of protons well exceeds the photopion production energy threshold at $z=0$.

The maximum proton energy dependence of the flux intensity of neutrinos is effectively determined by Eq.~(\ref{eq:optical_depth_proton}) in this formulation. In the case of the calculations for the high-redshift universe, the evolution of the CMB temperatures strongly affects the association of the Mandelstam variable through the kinematics of a collision. Equation~(\ref{eq:optical_depth_proton}) employs the rectangular $\Delta$-resonance approximations, which serve to replace the integral of the Madelstam variable. This substitution is made in order to simplify the mathematical calculations and enhance intuitive understanding of the formulation. The more accurate numerical integrals result in a shift in the peak intensity.  The degree of the systematic uncertainty in the peak intensity resulting from this simple analytical 
approximation is illustrated as the belt in Fig.~1.

An additional approximation that results in an underestimation of the intensity of neutrinos is the neglect of the contribution of the direct pion production channel in the photopion production. In the energy region below the $\Delta$-resonance energy, the interactions via the direct production are predominant. This process effectively reduces the threshold energy required for protons to produce charged pions. The \textsc{crp}{\small{ropa}} simulation package incorporates this channel through the implementation of the \textsc{sophia} Monte Carlo program~\cite{Mucke:1999yb}, while the analytical formulation does not consider it. In conventional cosmogenic neutrino flux calculations, where the proton spectrum is well extended above the photopion production energy threshold, the effect is subdominant. However, in the present scenario, where the maximum energy of protons emitted from a source is well below the threshold energy {\it in the present universe}, the moderate decrease of the effective energy threshold due to the direct pion production leads to a substantial enhancement of the neutrino production. Consequently, the quasi-calorimetric condition $\tau_0=1/\overline{x_\Delta}$, which is solely based on the $\Delta$(1232)-resonance approximation, inevitably underestimates the resultant cosmogenic neutrino flux in the LRD scenario. This factor is responsible for the apparent differences shown in Fig.~1, which has been confirmed by switching off the direct production channel in \textsc{crp}{\small{ropa}}.

\twocolumngrid
\bibliography{syoshida}

\end{document}